\documentclass[12pt,portuguese]{article}

\usepackage{babel}

\begin{document}

\title{\textbf{Um problema de tr\^{e}s corpos analiticamente sol\'uvel}\\
\vspace{0.3cm} (An analytically solvable three-body problem)}
\date{}
\author{Elysandra Figuer\^edo\thanks{%
E-mail: lys@urania.iagusp.usp.br} (1) e Antonio S. de Castro\thanks{%
E-mail: castro@feg.unesp.br} (2) \\
\\
(1) USP - Instituto Astron\^omico e Geof\'{\i}sico\\
Departamento de Astronomia\\
Caixa Postal 3386\\
01060-970 S\~ao Paulo SP\\
(2) UNESP - Campus de Guaratinguet\'a\\
Departamento de F\'{\i}sica e Qu\'{\i}mica\\
Caixa Postal 205\\
12500-000 Guaratinguet\'a SP}
\maketitle

\begin{abstract}
Analisamos um problema de tr\^{e}s corpos interagindo mutuamente via for\c{c}%
as harm\^{o}nicas no contexto do formalismo newtoniano. Uma solu\c{c}\~{a}o
anal\'{\i }tica exata para este problema \'{e} encontrada por meio de uma
abordagem did\'{a}tica e os caminhos para a an\'{a}lise do problema de $N $
corpos s\~{a}o indicados.

The problem of three particles interacting through harmonic forces is
discussed within the  Newtonian formalism. By means of a didactic approach,
an exact analytical solution is found, and ways to extend it to the $N$-body
case are pointed out.
\end{abstract}


\section{\textbf{Introdu\c{c}\~{a}o}}

Apesar dos esfor\c{c}os dos f\'{\i }sicos e matem\'{a}ticos por mais que
dois s\'{e}culos de pesquisa, o problema geral de $N$ corpos interagindo
mutuamente e movendo-se de acordo com as leis de Newton, para $N>2$, nunca
foi resolvido exatamente. O problema de dois corpos sujeito a for\c{c}as que
dependem do vetor posi\c{c}\~{a}o relativa pode ser reduzido a dois
problemas de um corpo, um dos quais descreve o movimento do centro de massa
e o outro o movimento re\-la\-ti\-vo. $N=3$ \'{e} o menor valor de $N$ que
torna o 
problema de $N$ corpos insol\'{u}vel no caso geral. Contudo, sob suposi\c{c}%
\~{o}es especiais a respeito do tipo de movimento e intera\c{c}\~{a}o, solu%
\c{c}\~{o}es anal\'{\i }ticas para o problema de $N$ corpos podem ser
encontradas.

No caso do problema de tr\^{e}s corpos com intera\c{c}\~{o}es gravitacionais
algumas solu\c{c}\~{o}es especiais s\~{a}o normalmente apresentadas nos
livros texto de mec\^{a}nica cl\'{a}ssica. No chamado problema restrito de
tr\^{e}s corpos, dois corpos pesados movem-se em torno do centro de massa
comum enquanto um terceiro corpo leve move-se no mesmo plano \cite{mar}-\cite
{hes}. No chamado caso de Lagrange os tr\^{e}s corpos est\~{a}o durante todo
o movimento sobre os v\'{e}rtices de um tri\^{a}ngulo equil\'{a}tero, que
gira em torno de um eixo perpendicular ao plano dos corpos enquanto troca de
tamanho \cite{mar2}-\cite{som}. Existe ainda uma outra solu\c{c}\~{a}o
especial para o problema de tr\^{e}s corpos interagindo gravitacionalmente
conhecida como caso de Euler. Neste \'{u}ltimo caso os corpos movem-se ao
longo da mesma linha reta durante todo o movimento \cite{sym3}-\cite{som2}.
Uma outra solu\c{c}\~{a}o especial \'{e} essa de $N$ corpos de massas
similares sujeita a for\c{c}as similares movendo-se sobre os v\'{e}rtices de
um pol\'{\i }gono regular de $N$ lados \cite{som3}. Todos estes movimentos
especiais s\~{a}o de grande import\^{a}ncia pedag\'{o}gica tendo em vista
que eles s\~{a}o solu\c{c}\~{o}es de um problema insol\'{u}vel no caso
geral. Contudo, a re\-so\-lu\-\c{c}\~{a}o do caso de Lagrange, como
apresentada pelos autores de livros-texto, recorre a um sistema de
coordenadas em rota\c{c}\~{a}o requerendo desta forma um c\'{a}lculo extenso
e elaborado, e conseq\"{u}entemente a um enfraquecimento da atratividade
pedag\'{o}gica.

Recentemente o caso de Lagrange foi apresentado de modo alternativo e mais
geral dentro do formalismo newtoniano, permitindo que ele possa ser
facilmente abordado imediatamente depois da apresenta\c{c}\~{a}o do problema
de dois corpos \cite{asc}. Nesse trabalho solu\c{c}\~{o}es de tri\^{a}ngulo
equil\'{a}tero foram obtidas para intera\c{c}\~{o}es que v\~{a}o al\'{e}m do
caso gravitacional. Ess\'{e}n, em um trabalho recente \cite{ess}
(hom\^{o}nimo a esse da Ref. \cite{asc}), abordou o mesmo problema usando o
formalismo lagrangiano, restringindo-se ao caso gravitacional e apresentando
uma extens\~{a}o ao problema de $N$ corpos. Encorajado pelos resultado
obtidos em \cite{ess} e seguindo os mesmos passos da Ref. \cite{asc}, foi
realizada uma extens\~{a}o do caso de Lagrange para um sistema de $N$ corpos 
\cite{asc2} buscando uma solu\c{c}\~{a}o para a qual a for\c{c}a sobre cada
um dos $N$ corpos est\'{a} na dire\c{c}\~{a}o do centro de massa do sistema
(a mesma imposi\c{c}\~{a}o j\'{a} u\-sa\-da em \cite{asc} e \cite{ess}).
Isto realmente acontece para intera\c{c}\~{o}es gravitacionais porque as for%
\c{c}as s\~{a}o proporcionais as massas dos corpos, mas tamb\'{e}m pode
acontecer para outras esp\'{e}cies de intera\c{c}\~{o}es desde que certas
condi\c{c}\~{o}es sejam satisfeitas pelas massas dos corpos e constantes de
for\c{c}a. Como subproduto obtivemos que os $N$ corpos est\~{a}o sobre os
v\'{e}rtices de uma figura geom\'{e}trica regular durante todo o movimento.
Para o caso de intera\c{c}\~{o}es harm\^{o}nicas chegou-se a conclus\~{a}o
que o problema de $N$ corpos reduz-se a $N$ problemas de um corpo, portanto
os $N$ corpos movem-se independentemente, n\~{a}o necessitando estarem sobre
os v\'{e}rtices de uma figura geom\'{e}trica regular. A condi\c{c}\~{a}o
envolvendo as massas dos corpos e as constantes de for\c{c}a para as for\c{c}%
as harm\^{o}nicas obtida em \cite{asc} e \cite{asc2} tem a mesma forma que
essa j\'{a} obtida em um trabalho anterior (usando o formalismo
lagrangiano): condi\c{c}\~{a}o necess\'{a}ria para as coordenadas de Jacobi
conduzir \`{a} separa\c{c}\~{a}o de vari\'{a}veis no problema de tr\^{e}s
corpos com intera\c{c}\~{o}es harm\^{o}nicas \cite{asc3}.

No presente trabalho apresentamos o problema j\'{a} abordado na Ref. \cite
{asc3} mas desta feita usamos o formalismo newtoniano. Veremos que este
problema especial de tr\^{e}s corpos pode ser abordado com facilidade
suficiente para que possa ser apresentado aos estudantes de ci\^{e}ncias
exatas ainda no primeiro semestre dos cursos de gradua\c{c}\~{a}o. A
import\^{a}ncia deste problema espec\'{\i }fico n\~{a}o \'{e} apenas
pedag\'{o}gica tendo em vista que o problema de tr\^{e}s corpos, interagindo
mutuamente via for\c{c}as harm\^{o}nicas, tem sido usado no c\'{a}lculo da
espectroscopia de b\'{a}rions no modelo de quarks \cite{fla}-\cite{kal}. Come%
\c{c}aremos revendo o problema geral de dois dois corpos e em seguida o
problema de tr\^{e}s corpos com intera\c{c}\~{o}es harm\^{o}nicas sem
incluir \textit{a priori} restri\c{c}\~{o}es sobre as massas dos corpos e
constantes de for\c{c}a.

\section{\textbf{O problema de dois corpos}}

As equa\c{c}\~{o}es de movimento para dois corpos de massas $m_{1}$ e $m_{2}$%
, localizadas pelos vetores posi\c{c}\~{a}o $\vec{r}_{1}$ e $\vec{r}_{2}$,
respectivamente, podem ser escritas como

\begin{equation}
\vec{F}_{1}\left( \vec{r}_{1},\vec{r}_{2}\right) =m_{1}\ddot{\vec{r}_{1}}
\label{eq1}
\end{equation}

\begin{equation}
\vec{F}_{2}\left( \vec{r}_{1},\vec{r}_{2}\right) =m_{2}\ddot{\vec{r}_{2}}
\label{eq2}
\end{equation}

\noindent onde $\vec{F}_{1}$ ($\vec{F}_{2}$) \'{e} a for\c{c}a que o corpo 2
(1) exerce sobre o corpo 1 (2), e cada ponto sobre os vetores posi\c{c}\~ao $%
\vec{r}_{1}$ e $\vec{r}_{2}$ denota uma derivada temporal. Observe que as for%
\c{c}as dependem dos vetores posi\c{c}\~{a}o dos corpos de forma que as equa%
\c{c}\~{o}es (\ref{eq1}) e (\ref{eq2}) s\~{a}o equa\c{c}\~{o}es diferenciais
acopladas. Introduzindo o vetor posi\c{c}\~{a}o do centro de massa e o vetor
coordenada relativa:

\begin{equation}
\vec{R}=\frac{m_{1}\vec{r}_{1}+m_{2}\vec{r}_{2}}{m_{1}+m_{2}}  \label{eq3}
\end{equation}

\begin{equation}
\vec{r}=\vec{r}_{1}-\vec{r}_{2}  \label{eq4}
\end{equation}

\noindent os vetores $\vec{r}_{1}$ e $\vec{r}_{2}$ podem ser escritos como

\begin{equation}
\vec{r}_{1}=\vec{R}+\frac{m_{2}}{m_{1}+m_{2}}\vec{r}  \label{eq5}
\end{equation}

\begin{equation}
\vec{r}_{2}=\vec{R}-\frac{m_{1}}{m_{1}+m_{2}}\vec{r}  \label{eq6}
\end{equation}

\noindent Quando (\ref{eq5}) e (\ref{eq6}) s\~{a}o introduzidos em (\ref{eq1}%
) e (\ref{eq2}) e considerando a forma fraca da terceira lei de Newton
(n\~{a}o se exigindo que as for\c{c}as tenham a mesma dire\c{c}\~{a}o da
linha que une os dois corpos), resulta que

\begin{equation}
M\ddot{\vec{R}}=\vec{0}  \label{eq7}
\end{equation}

\begin{equation}
\vec{F}\left( \vec{r}_{1},\vec{r}_{2}\right) =\mu \vec{r}  \label{eq8}
\end{equation}

\noindent onde $\vec{F}\left( \vec{r}_{1},\vec{r}_{2}\right) =\vec{F}%
_{1}\left( \vec{r}_{1},\vec{r}_{2}\right) $, $M$ \'e a massa do sistema e

\begin{equation}
\frac{1}{\mu }=\frac{1}{m_{1}}+\frac{1}{m_{2}}  \label{eq9}
\end{equation}

\noindent Considerando ainda que as for\c{c}as dependem das posi\c{c}\~{o}es
dos corpos apenas pelo vetor posi\c{c}\~{a}o relativa, chegamos finalmente a
conclus\~{a}o que

\begin{equation}
\vec{F}\left( \vec{r}\right) =\mu \ddot{\vec{r}}  \label{eq10}
\end{equation}

\noindent Os resultados (\ref{eq7}) e (\ref{eq10}) mostram que o problema de
dois corpos sob intera\c{c}\~{a}o m\'{u}tua foi finalmente reduzido a dois
problemas de um corpo. Um dos pro\-ble\-mas \'{e} aquele de um corpo livre
de massa igual a massa total do sistema localizado pelo vetor posi\c{c}%
\~{a}o do centro de massa. O outro problema \'{e} aquele de um corpo de
massa ${\mu }$, chamada de massa reduzida, localizado pelo vetor posi\c{c}%
\~{a}o relativa. Toda a dificuldade da solu\c{c}\~{a}o do problema de dois
corpos reside agora na busca de solu\c{c}\~{a}o deste \'{u}ltimo problema de
um corpo.

\section{\textbf{Um problema de tr\^{e}s corpos}}

As equa\c{c}\~{o}es de movimento para tr\^{e}s corpos de massas $m_{i}$ ($%
i=1...3$), localizados pelos vetores posi\c{c}\~{a}o $\vec{r}_{i}$ ($i=1...3$%
), respectivamente, podem ser escritas como

\begin{equation}
\vec{F}_{1}\left( \vec{r}_{1},\vec{r}_{2},\vec{r}_{3}\right) =m_{1}\ddot{%
\vec{r}_{1}}  \label{eq11}
\end{equation}

\begin{equation}
\vec{F}_{2}\left( \vec{r}_{1},\vec{r}_{2},\vec{r}_{3}\right) =m_{2}\ddot{%
\vec{r}_{2}}  \label{eq12}
\end{equation}

\begin{equation}
\vec{F}_{3}\left( \vec{r}_{1},\vec{r}_{2},\vec{r}_{3}\right) =m_{3}\ddot{%
\vec{r}_{3}}  \label{eq13}
\end{equation}

\noindent Supomos que as intera\c{c}\~{o}es m\'{u}tuas s\~{a}o intera\c{c}%
\~{o}es entre pares de corpos e que as for\c{c}as s\~{a}o diretamente
proporcionais \`{a} coordenada relativa (for\c{c}as harm\^{o}nicas):

\begin{equation}
\vec{F}_{1}=-K_{12}\left( \vec{r}_{1}-\vec{r}_{2}\right) -K_{13}\left( \vec{r%
}_{1}-\vec{r_{3}}\right)  \label{eq14}
\end{equation}

\begin{equation}
\vec{F}_{2}=-K_{21}\left( \vec{r}_{2}-\vec{r}_{1}\right) -K_{23}\left( \vec{r%
}_{2}-\vec{r}_{3}\right)  \label{eq15}
\end{equation}

\begin{equation}
\vec{F}_{3}=-K_{31}\left( \vec{r}_{3}-\vec{r}_{1}\right) -K_{32}\left( \vec{r%
}_{3}-\vec{r}_{2}\right)  \label{eq16}
\end{equation}

\noindent $K_{ij}>0$ s\~{a}o as constantes de for\c{c}a obedecendo \`{a} rela%
\c{c}\~{a}o $K_{ij}=K_{ji}$, em conformidade com a terceira lei de Newton na
forma fraca. Observa-se aqui que a forma forte da terceira lei de Newton,
estabelecendo que as for\c{c}as m\'{u}tuas al\'{e}m de terem os mesmos
m\'{o}dulos e sentidos opostos t\^{e}m que ter a dire\c{c}\~{a}o da linha
que une os corpos, \'{e} automaticamente obedecida. Aqui tamb\'{e}m, mais
explicitamente que no caso de dois corpos tratado na se\c{c}\~{a}o anterior,
\'{e} visto que as equa\c{c}\~{o}es de movimento s\~{a}o equa\c{c}\~{o}es
diferenciais acopladas. As coordenadas de Jacobi s\~{a}o definidas como

\begin{equation}
\vec{R}=\frac{m_{1}\vec{r}_{1}+m_{2}\vec{r}_{2}+m_{3}\vec{r}_{3}}{%
m_{1}+m_{2}+m_{3}}  \label{eq17}
\end{equation}

\begin{equation}
\vec{\rho}=\vec{r}_{1}-\vec{r}_{2}  \label{eq18}
\end{equation}

\begin{equation}
\vec{\lambda}=\vec{r}_{3}-\frac{m_{1}\vec{r}_{1}+m_{2}\vec{r}_{2}}{%
m_{1}+m_{2}}  \label{eq19}
\end{equation}

\noindent onde $\vec{R}$ \'{e} a coordenada do centro de massa do sistema de
tr\^{e}s corpos, $\vec{\rho}$ \'{e} a coordenada do corpo 1 relativa ao
corpo 2, e $\vec{\lambda}$ \'{e} a coordenada do corpo 3 relativa ao centro
de massa dos corpos 1 e 2. Em termos das coordenadas de Jacobi os vetores
posi\c{c}\~{a}o $\vec{r}_{i}$ podem ser escritos como

\begin{equation}
\vec{r}_{1}=\vec{R}-\frac{m_{3}}{M}\vec{\lambda}+\frac{m_{2}}{M_{12}}\vec{%
\rho}  \label{eq20}
\end{equation}

\begin{equation}
\vec{r}_{2}=\vec{R}-\frac{m_{3}}{M}\vec{\lambda}-\frac{m_{1}}{M_{12}}\vec{%
\rho}  \label{eq21}
\end{equation}

\begin{equation}
\vec{r}_{3}=\vec{R}+\frac{M_{12}}{M}\vec{\lambda}  \label{eq22}
\end{equation}

\noindent onde $M$ \'{e} a massa do sistema e $M_{12}=m_{1}+m_{2}$ \'{e} a
massa do subsistema constitu\'{\i}do pelos corpos 1 e 2. Quando (\ref{eq14}%
)-(\ref{eq16}) e (\ref{eq20})-(\ref{eq22}) s\~{a}o introduzidas em (\ref
{eq11})-(\ref{eq13}) resulta que

\begin{equation}
\ddot{\vec{R}}=\vec{0}  \label{eq23}
\end{equation}

\begin{equation}
\ddot{\vec{\rho}}+\omega _{1}^{2}\,\vec{\rho}=\frac{\Gamma }{M_{1}}\vec{%
\lambda}  \label{eq24}
\end{equation}

\begin{equation}
\ddot{\vec{\lambda}}+\omega _{2}^{2}\,\vec{\lambda}=\frac{\Gamma }{M_{2}}%
\vec{\rho}  \label{eq25}
\end{equation}

\noindent onde

\begin{equation}
\frac{1}{M_{1}}=\frac{1}{m_{1}}+\frac{1}{m_{2}}  \label{eq26}
\end{equation}

\begin{equation}
\frac{1}{M_{2}}=\frac{1}{m_{1}+m_{2}}+\frac{1}{m_{3}}  \label{eq27}
\end{equation}

\begin{equation}
\omega _{1}^{2}=\frac{1}{M_{1}}\left( K_{12}+\frac{%
K_{13}m_{2}^{2}+K_{23}m_{1}^{2}}{M_{12}^{2}}\right)  \label{eq28}
\end{equation}

\begin{equation}
\omega _{2}^{2}=\frac{1}{M_{2}}\left( K_{13}+K_{23}\right)  \label{eq29}
\end{equation}

\begin{equation}
\Gamma =\frac{K_{13}m_{2}-K_{23}m_{1}}{M_{12}}  \label{eq30}
\end{equation}

\noindent Pode-se observar destes resultados que as coordenadas de Jacobi
reduziram este problema de tr\^{e}s corpos ao movimento livre do centro de
massa (\ref{eq23}), conseq\"{u}\^{e}ncia da aus\^{e}ncia de for\c{c}as
externas, e ao movimento de dois osciladores harm\^{o}nicos acoplados pela
constante $\Gamma $, que anula-se somente quando $K_{13}m_{2}=K_{23}m_{1}$.

Para desacoplar as equa\c{c}\~{o}es de movimento no caso geral de massas e
constantes de for\c{c}a devemos recorrer a um outro conjunto de coordenadas.
Vamos considerar uma transforma\c{c}\~{a}o de coordenadas que \'{e} uma
mistura de uma transforma\c{c}\~{a}o de escala e uma rota\c{c}\~{a}o \cite
{asc3}, \cite{dut}, definida por

\begin{equation}
\vec{\rho}=\left( \frac{M_{E}}{M_{1}}\right) ^{1/2}\cos (\phi )\,\vec{y}%
_{1}-\left( \frac{M_{E}}{M_{1}}\right) ^{1/2}\sin (\phi )\,\vec{y}_{2}
\label{eq31}
\end{equation}

\begin{equation}
\vec{\lambda}=\left( \frac{M_{E}}{M_{2}}\right) ^{1/2}\sin (\phi )\,\vec{y}%
_{1}+\left( \frac{M_{E}}{M_{2}}\right) ^{1/2}\cos (\phi )\,\vec{y}_{2}
\label{eq32}
\end{equation}

\noindent \noindent onde $M_{E}$ \'{e} um par\^{a}metro com dimens\~{a}o de
massa e $\phi $ \'{e} um par\^{a}metro de rota\c{c}\~{a}o. A princ\'{\i}pio
os par\^{a}metros $M_{E}$ e $\phi $ s\~{a}o arbitr\'{a}rios. Inserindo (\ref
{eq31}) e (\ref{eq32}) em (\ref{eq24}) e (\ref{eq25}) encontramos novas equa%
\c{c}\~{o}es diferenciais acopladas para as novas coordenadas:

\begin{equation}
\ddot{\vec{y}}_{1}+\alpha ^{2}\,\vec{y}_{1}=\gamma \vec{y}_{2}  \label{eq33}
\end{equation}

\begin{equation}
\ddot{\vec{y}}_{2}+\beta ^{2}\vec{y}_{2}=\gamma \vec{y}_{1}  \label{eq34}
\end{equation}

\noindent onde

\begin{equation}
\alpha^2 =\omega _{1}^{2}\cos ^{2}(\phi )+\omega _{2}^{2}\sin ^{2}(\phi )-%
\frac{\Gamma }{\left( M_{1}M_{2}\right) ^{1/2}}\sin \left( 2\phi \right)
\label{eq35}
\end{equation}

\begin{equation}
\beta^2 =\omega _{1}^{2}\sin ^{2}(\phi )+\omega _{2}^{2}\cos ^{2}(\phi )+%
\frac{\Gamma }{\left( M_{1}M_{2}\right) ^{1/2}}\sin \left( 2\phi \right)
\label{eq36}
\end{equation}

\begin{equation}
\gamma =\frac{1}{2}\left( \omega _{1}^{2}-\omega _{2}^{2}\right) \sin \left(
2\phi \right) +\frac{\Gamma }{\left( M_{1}M_{2}\right) ^{1/2}}\cos \left(
2\phi \right)  \label{eq37}
\end{equation}

\noindent A elimina\c{c}\~{a}o do acoplamento entre as equa\c{c}\~{o}es
diferenciais (\ref{eq33}) e (\ref{eq34}) pode ser obtida se pu\-der\-mos
tomar proveito da ar\-bi\-tra\-ri\-e\-da\-de do va\-lor do par\^{a}metro $%
\phi $ impondo que $\gamma =0$. Isto realmente acontece quando

\begin{equation}
\tan \left( 2\phi \right) =-\frac{2\Gamma }{\left( \omega _{1}^{2}-\omega
_{2}^{2}\right) \left( M_{1}M_{2}\right) ^{1/2}}  \label{eq38}
\end{equation}
Com $\phi $ dado por (\ref{eq38}) finalmente obtemos as seguintes equa\c{c}%
\~{o}es diferenciais desacopladas:

\begin{equation}
\ddot{\vec{y}}_{1}+\Omega _{1}^{2}\,\vec{y}_{1}=\vec{0}  \label{eq39}
\end{equation}

\begin{equation}
\ddot{\vec{y}}_{2}+\Omega _{2}^{2}\,\vec{y}_{2}=\vec{0}  \label{eq40}
\end{equation}

\noindent onde $\Omega _{1}=\Omega _{+},\Omega _{2}=\Omega _{-}$ e

\begin{equation}
\Omega _{\pm }=\frac{1}{2}\left( \omega _{1}^{2}+\omega _{2}^{2}\right) \pm 
\frac{1}{2}\left[ \left( \omega _{1}^{2}-\omega _{2}^{2}\right) ^{2}+\frac{%
4\Gamma ^{2}}{M_{1}M_{2}}\right] ^{1/2}  \label{eq41}
\end{equation}

\noindent As equa\c{c}\~{o}es diferenciais (\ref{eq39}) e (\ref{eq40})
descrevem o movimento de dois osciladores harm\^{o}nicos desacoplados, cujas
solu\c{c}\~{o}es $\vec{y}_{1}(t)$ e $\vec{y}_{2}(t)$ s\~{a}o bem
co\-nhe\-ci\-das. Usando as transforma\c{c}\~{o}es (\ref{eq31})-(\ref{eq32}%
), e em seguida (\ref{eq20})-(\ref{eq22}), obteremos as solu\c{c}\~{o}es
anal\'{\i }ticas para $\vec{r}_{i}(t)$. Desnecess\'{a}rio mencionar a solu\c{%
c}\~{a}o $\vec{R}(t)$ da equa\c{c}\~{a}o diferencial (\ref{eq23}).

\section{Conclus\~{a}o}

Neste artigo abordamos o problema de tr\^{e}s corpos interagindo mutuamente
via for\c{c}as harm\^{o}nicas. O problema poderia ter sido abordado com
muito maior simplicidade se consider\'{a}ssemos desde o in\'{\i }cio corpos
com massas e constantes de for\c{c}as similares no sistema de refer\^{e}ncia
do centro de massa ($\vec{R}=\vec{0}$). Encontrar\'{\i }amos ent\~{a}o que
as coordenadas de Jacobi teriam tido o \^{e}xito procurado. Optamos por
n\~{a}o impor tal severa restri\c{c}\~{a}o \textit{a priori,} e mostramos
que as coordenadas de Jacobi n\~{a}o desacoplam o problema no caso geral mas
somente quando $K_{13}m_{2}=K_{23}m_{1}$, um resultado que contrasta com
esse encontrado na literatura \cite{fla}-\cite{kal}, onde considera-se que o
desacoplamento ocorre para massas gen\'{e}ricas e constantes de for\c{c}a
similares. A assimetria apresentada por essa condi\c{c}\~{a}o \textit{sine
qua non} para o desacoplamento surge em decorr\^{e}ncia da assimetria na
defini\c{c}\~{a}o das coordenadas de Jacobi $\vec{\rho}$ e $\vec{\lambda}$.
Em geral, as coordenadas de Jacobi conduzem o problema de tr\^{e}s corpos
interagindo mutuamente via for\c{c}as harm\^{o}nicas ao problema do
movimento livre do centro de massa mais dois osciladores harm\^{o}nicos
acoplados, ainda que as constantes de for\c{c}a sejam id\^{e}nticas.
Tamb\'{e}m mostramos que existe um outro sistema de coordenadas que conduz
\`{a} separabilidade no caso geral. Restringimos nossa aten\c{c}\~{a}o ao
caso de tr\^{e}s corpos mas pode-se verificar que extens\~{o}es para o caso
de $N$ corpos s\~{a}o tamb\'{e}m pass\'{\i }veis de solu\c{c}\~{o}es anal\'{%
\i }ticas. Para o problema de quatro corpos, em particular, precisar\'{\i %
}amos redefinir o vetor posi\c{c}\~{a}o do centro de massa e acrescentar uma
nova coordenada de Jacobi, essa nova coordenada descrevendo a posi\c{c}%
\~{a}o do corpo 4 relativa ao centro de massa dos corpos 1, 2 e 3. Quando as
massas e as constantes de for\c{c}a s\~{a}o similares as coordenadas de
Jacobi \textit{per se} conduzem \`{a} separabilidade deste problema de
quatro corpos, em caso contr\'{a}rio deveremos recorrer a um sistema de
coordenadas adicional, misturando transforma\c{c}\~{a}o de escala e rota\c{c}%
\~{a}o no espa\c{c}o tridimensional, quando ent\~{a}o deveremos lidar com
tr\^{e}s par\^{a}metros de rota\c{c}\~{a}o relacionados com os \^{a}ngulos
de Euler. Estas tarefas s\~{a}o deixadas para os leitores.

\section{Agradecimentos}

Os autores s\~ao gratos \`a FAPESP pelo apoio financeiro.

\end{document}